# Auto-encoder based Model for High-dimensional Imbalanced Industrial Data

Chao Zhang[1]; Sthitie Bom[1,*]

[1] Seagate Technology, Bloomington, MN 55435, USA
sthitie.e.bom@seagate.com

**Abstract.** With the proliferation of IoT devices, the distributed control systems are now capturing and processing more sensors at higher frequency than ever before. These new data, due to their volume and novelty, cannot be effectively consumed without the help of data-driven techniques. Deep learning is emerging as a promising technique to analyze these data, particularly in soft sensor modeling. The strong representational capabilities of complex data and the flexibility it offers from an architectural perspective make it a topic of active applied research in industrial settings. However, the successful applications of deep learning in soft sensing are still not widely integrated in factory control systems, because most of the research on soft sensing do not have access to large scale industrial data which are varied, noisy and incomplete. The results published in most research papers are therefore not easily reproduced when applied to the variety of data in industrial settings. Here we provide manufacturing data sets that are much larger and more complex than public open soft sensor data. Moreover, the data sets are from Seagate factories on active service with only necessary anonymization, so that they reflect the complex and noisy nature of real-world data. We introduce a variance weighted multi-headed auto-encoder classification model that fits well into the high-dimensional and highly imbalanced data. Besides the use of weighting or sampling methods to handle the highly imbalanced data, the model also simultaneously predicts multiple outputs by exploiting output-supervised representation learning and multi-task weighting.

## 1 INTRODUCTION

### 1.1 Background

There has been considerable interest and investment in the exploitation of smart sensors among government, academia and industry in the last decade.

Data-driven soft sensing models in the past few years have generated active research and investment due their efficient handling of complex data relationships and reduced reliance on first principles. These models are primarily based on the data captured about the processes and are more flexible and more easily deployed in real industrial plants due to their minimal reliance on domain expert's knowledge.

These methods especially shine in modern industrial processes which demand sophisticated monitoring and control mechanisms that can efficiently handle the complexities of time, space and non-linearity. Most state-of-the-art industrial plants have distributed control systems that collect large amounts of sensor data, which can be mined to extract relevant information critical to process control.

Soft sensor models are typically developed to effectively capture and control the quality variations in different processes run by the equipment or alternatively, variations in the equipment itself. In industrial data, some of the typical challenges are the non-linear complexities of the process variables, the sampling discrepancy between the process variables and the quality outcomes, and the lack of labelled data. These challenges are compounded by the degradation of the models caused by the inherent fluctuation in the raw material properties and the variations in the product mix being processed by the equipment.

The attraction of data-driven soft sensor modeling is that it offers reliable and economical alternatives to rule-based, first principles-driven expensive measuring sensors, thereby enabling effective feedback, and control strategies for process monitoring [1].

A variety of deep learning techniques have been proposed over the last years in soft sensor application, including stacked auto-encoders, deep belief networks, and flavors of recurrent and convolutional neural nets. The interest is motivated by promising empirical results of these deep learning techniques, which show improved representation ability, ability to exploit the unlabeled data to improve performance, and potential for non-linear feature extraction.

While the body of research around soft sensing has been non-trivial and has contributed well to the increased research interest in data-driven soft sensing, we still see a gap between laboratory outcomes and industrial deployments of these soft sensors [15].

One of the reasons for this gap is the lack of real-world data that these models can sufficiently be exercised upon. For example, most of the soft sensors related work cited in this article, use datasets such as the benchmark TE (Tennessee Eastman) and the Debutanizer column. These have relatively low volume and dimensions (number of samples up to 30K and dimensions up to 52). A real distributed control system in an industrial plant, on the other hand, monitors thousands of operations generating tens of thousands of sensors from hundreds of equipment capturing millions of observations per day. Additionally, there are inherent data imbalance problem that none of these studies capture and address. From a practical implementation perspective, we believe that there is also a need to have an architecture that can into not only learn the interaction of all these sensors within a tool but also be able to simultaneously predict operations down the manufacturing line, so proactive process monitoring and controls can be established.

This paper proposes a scalable soft sensor application that is motivated by the above-mentioned current gap in the variety of process data for research and the speed of deployment and generalizability for broad coverage in industrial process control. We propose a variance weighted multi-headed classification model that simultaneously predicts multiple outputs by exploiting output-supervised representation learning and multi-task weighting.

Our contributions are:
1. A generalized imbalance-weighted algorithm that uses sensor data captured from the process tools to predict multiple outcomes measured by multiple measurement tools
2. Open release of the preprocessed real-world data to advance further research in soft sensing using a diverse family of semiconductor process manufacturing tools.

In the following parts of this paper, we summarize some related work in part II, and autoencoders are introduced in part III, followed by the proposed variance weighted multi-headed classification model. Part V presents a case study on wafer manufacturing data from Seagate that is carried out to show the performance of the proposed model. Results are discussed in part VI.

## 1.2 Related Work

Over the last decades, most deep learning-based soft sensor modeling have used variants of auto-encoders as a pre-training approach. This was motivated by the work done in [2], which demonstrated with extensive experiments that unsupervised pre-training provides good marginal distribution and captures more intricate dependencies between parameters. In other words, it acts like a pre-conditioner, providing a more suitable runway for future supervised training.

However, the challenge is that even though deep learning techniques extract high-level features from the data, these features may not always be relevant to discriminating the specific problem at hand, for example, detecting the quality of a product or a fault in a piece of equipment. To address this, several researchers have tried to inject relevant output variables during pre-training to guide the training towards learning features that are more discriminating of the classes and outputs of interest.

An automated key feature learning method was presented by [9] for nonlinear process monitoring, which applies the k- nearest neighbor rule to find the nearest samples in the training set. The autoencoder is utilized to convert the input space into a feature (a de-noised reconstructed X) and residual (input X – reconstructed X) space. Control limits are derived by kernel density estimation, so pertinent fault detection alarms can be triggered when the appropriate thresholds are crossed.

[6] extract more discriminative features by proposing an autoencoder-inspired unsupervised feature selection method that can jointly learn to reconstruct its inputs and the importance weights of each feature by simultaneously minimizing reconstruction error and group sparsity regularization.

[12] leverage the unsupervised stacked auto-encoder to replace the random weight initialization strategy adopted in deep LSTM recurrent networks. Their proposed LSTM-SAE uses unsupervised pre-trained LSTMs in an autoencoder fashion. Their results show that an unsupervised AE based pre-training approach improves the performance of deep LSTM using time series data and leads to faster convergence than other models.

Autoencoders are also implemented by [8] for feature extraction of process variables. Their solution implements a semi-supervised extreme learning machine based on manifold regularization in the last hidden layer of the deep autoencoder to obtain better prediction of process variables. In addition to exploiting a large number of unlabeled process data in the learning process, these extreme learning machines have the advantage of being computationally more efficient because they do not use resource-intensive back propagation.

As applied research continues in the industrial space for process monitoring, however, it is becoming clear that while unsupervised pre-training techniques do a good job of extracting high level features from the data, these features do not always represent the specific metrics of interest such as quality of a product or fault in an equipment unit etc. This inhibits their widespread implementation in the actual process control. To address this gap of making representations learn statistical structures of process interest, there have been multiple efforts at making representation more relevant to outputs of interest by injecting relevancy into the pre-training process.

[3] introduced variable wise feature selection, where the process variables were weighted by their correlations with the output variable so as to extract high-level output-related features. This technique used the Pearson correlation, which is a linear correlation measure, and the features extracted may not fully represent the non-linear relationships. [4] proposed a high-level feature learning by introducing an output-relevant loss function into the pre-training process so that the network can simultaneously learn to reconstruct input features and the quality variable. They applied this to demonstrate higher performance in regression tasks. Information considered to be irrelevant was eliminated by [5] calculating the mutual information in each layer between the representation and output layer to reveal high-order relationships. Their results show that irrelevant representations are eliminated during the training of the subsequent layer of a stacked auto-encoder.

The idea of gated units is exploited by [7] to modulate the flow of information in a stacked autoencoder where each hidden layer's contribution is considered to predict the target variable. They hypothesize that there are important representations at even shallow layers and therefore each must be able to contribute to the final prediction. They also use an ensemble strategy to make more effective use of unlabeled samples.

[11] also attempt to make the latent structure more relevant to the outcome of interest by adding a distance penalty into the loss function, which enabled the auto-encoder to extract more suitable representations by increasing the inter-class separability of labeled data. They provide a mutual information-based mathematical interpretation to support the effectiveness of their proposed technique.

Other studies leverage convolutions and sequence modeling with variants of CNN and LSTMs. [13] explore multichannel CNNs to extract the correlations of different process variables regardless of their topological distance. [14] integrate the CNN and LSTM architectures to predict particulate matter concentration. Their proposed architecture includes a CNN to extract features in both spatial and temporal domains. LSTM is then used to analyze the extracted features and perform forecasting.

As can be seen from above, autoencoder based feature extraction and relevant feature learning using a combination of deep learning techniques such as stacked autoencoders, semi-supervised autoencoders, CNN, and LSTMs have been widely studied with varying levels of success in industrial data.

## 2 Methodology

### 2.1 Autoencoders

An autoencoder is a neural network that tries to learn the underlying latent variables by encoding its input. Raw input data $x$ is encoded into a low dimensional latent feature space $h$ with an encoder function. The decoder function then attempts to reconstruct the input to $\bar{x}$ using the latent space. The autoencoder is trained by minimizing the distance between the original input $x$ and its reconstructed counterpart $\bar{x}$. Suppose $x \in \mathbb{R}^{d_x}$ where $d_x$ is the dimension of the input data and $h \in \mathbb{R}^{d_h}$ where $\mathbb{R}^{d_h}$ is the dimension of the encoded vector. The encoder, decoder, and the loss functions in an autoencoder (AE) then can be expressed as follows:

$$h = f_{encoder}(W_{encoder} \cdot x + b_{encoder}) \qquad (1)$$
$$\bar{x} = f_{decoder}(W_{decoder} \cdot x + b_{decoder}) \qquad (2)$$
$$J_x(\theta_{AE}) = \frac{1}{N}\sum_{n=1}^{N} ||x - \bar{x}||^2 \qquad (3)$$

where $h$ is the encoded feature and $f_{encoder}(\cdot)$ and $f_{decoder}(\cdot)$ are non-linear activation functions that are used in the encoding and decoding process. $N$ is the total number of training samples. Autoencoders can be seen as special case of feedforward networks and may be trained with all the same techniques such as gradient descent following gradients derived by the back-propagation algorithm. Therefore, the parameters $\theta_{AE} = \{W_{encoder}, b_{encoder}, W_{decoder}, b_{decoder}\}$ in Equation 3 can be updated using gradient descent as follows:

$$\theta^* = arg \min_{\theta} Loss_{recon}(\theta_{AE}) \qquad (4)$$

Autoencoders are ideal for data which are characterized by high volume, and co-linearity and not fully understood non-linearities. By using a non-linear activation function such a sigmoid or a rectified linear unit (ReLU), the autoencoders can not only reduce the dimension space but also learn a more powerful generalization compared to PCA.

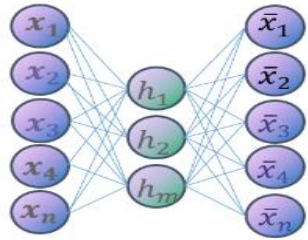

Fig. 1 A simple 3-layer autoencoder

However, one shallow network may not be able to fully represent the complex statistical structures of the input data. A common strategy is to greedily pretrain the deep autoencoders by training a stack of shallow autoencoders, thereby allowing the deep network to learn more abstract features hierarchically.

The self-supervised representational capabilities of the autoencoders can be exploited further to make them more relevant towards specific tasks such as predicting a failure of specific processes. This has been done by [4] with improved experimental results. They introduce a quality-based autoencoder (QAE) which attempts to reconstruct both the input data $x$ and the quality data $y$ simultaneously, which are denoted as $\bar{x}$ and $\bar{y}$. They add a prediction error for the target variable in the loss function of the traditional autoencoder in Equation 3. The parameter set $\theta_{AE}$ defined above is changed to $\theta_{QAE} = \{W_{encoder}, b_{encoder}, W_{decoder-x}, b_{decoder-x}, W_{decoder-y}, b_{decoder-y}\}$ which are jointly updated in the loss function as shown in Equation 5. As shown in Equation 5, the loss function then jointly minimizes the reconstruction loss and the prediction error between the true and the predicted label.

$$J(\theta_{QAE}) = \frac{1}{N}\sum_{n=1}^{N}(|| x - \bar{x} ||^2 + || y - \bar{y} ||^2) \qquad (5)$$

The main intuition here is that a combined loss function that minimizes the gap between true and predicted labels when added to the autoencoder reconstruction loss function, can guide the autoencoders towards reconstructing more specialized or relevant feature space.

This idea of guiding the unsupervised learning with a target variable can then be further developed to multi-outcome tasks. Consider, a learning task that maps the relationship between the sensors of one process machine in an industrial plant and the various outcomes as measured by multiple measurement machines down the line.

Instead of modeling each combination of process stage and measurement outcome step separately, we can model that process with all its measurement outcomes to jointly learn the outcome-relevant representation of each tool and outcomes of a specific family of measurement tools. At each process stage completion then, a classifier predicts class labels for not just the next step, but all the subsequent steps that the process would be measured at, thereby enabling continuous monitoring and proactive control of downstream processes.

### 2.2 Variance-Weighted Multi-Headed Quality-driven Autoencoder

While existing autoencoder based models have shown good performance on public data sets like debutanizer column data, their work does not consider the imbalance in the class labels, the units of the combined loss functions and may be prohibitive in environments where there are a large amount of process and output quality variables. To address these practical limitations, we propose a variance weighted multi-headed quality-driven auto-encoder (VWMHQAE) classification architecture which learns and predicts multiple quality-relevant features simultaneously. This is a combined method that predicts all measurement steps within the same model, thus much more efficient than individual methods.

We simplified the soft sensing problem into a classification problem because our main objective is to predict whether a wafer belongs to a failed outcome or a successful one. This is considered to be easier to interpret compared to a numerical output. At a mathematical level, a classification model is no different from regression except that the loss function is cross-entropy instead of mean-square-error. The formula of cross-entropy loss $L_{ce}$ is as below, with y as label and p as predicted probability:

$$L_{ce} = -y\log\hat{y} - (1-y)\log(1-\hat{y}) \qquad (6)$$

In a manufacturing line at any given processing step, there are usually a number of sensors being captured which can be used to monitor and analyze multiple quality attributes of that process. Typically, these quality measurements are performed across multiple measurement steps. To mimic this process, we designed a model with multiple heads, with each head denoting a measurement step. At each measurement step, specific properties of the wafer are measured to determine whether the properties pass the specification, which corresponds to a binary classification problem. As shown in Fig. 2A, several binary classification units are put together, to form a combined model to predict all the measurement steps at the same time.

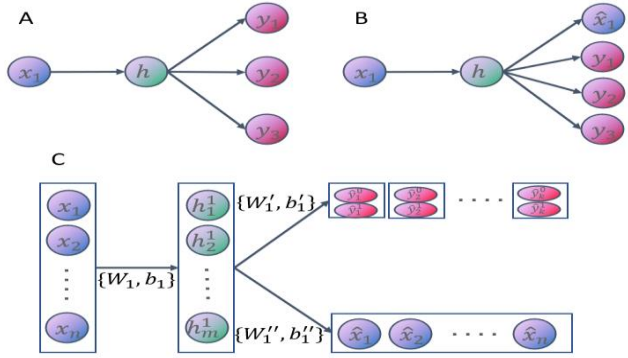

Fig. 2. A) Illustration of multi-headed model. B) Learning the reconstruction and target variable in the same time. C) multi-headed autoencoder model

To calculate the loss functions for all the binary classifications together, we added a weight to each class for each head to handle the imbalance from the source data. There are two terms in the Equation 7, one for the positive cases and one for the negative. The loss function for the entire classification problem is as below:

$$J_y(W,b) = \frac{1}{N} \sum_j^{N_h} \sum_i^{n_j} [-y_{i,j}^0 * log(\tilde{y}_{i,j}^0) * w_j^0 - y_{i,j}^1 * log(\tilde{y}_{i,j}^1) * w_j^1] \quad (7)$$

where N is the number of data points, $N_h$ is the number of heads, $n_j$ is the number of data points for the $j^{th}$ head, W, b are the model weights and bias, and $w_j^0, w_j^1$ are the class weights for negative and positive cases for the $j^{th}$ head. The class weights are calculated as following, with t indicating the negative and positive labels:

$$w_j^t = \frac{N}{2N_h * n_j^t} \quad (8)$$

Note that the $J_y$ in Equation 7 is not yet the final loss function for our model. To learn a representation of the input data and the features relating to the target variables in the same time, an auto-encoder based architecture is needed. This requires the objective function to be modified to include loss functions for both input reconstruction and target prediction. In our case, the reconstruction loss is an MSE loss for the input variables $J_x$ same as Equation 3, and the target prediction loss is $J_y$ as Equation 7. Fig. 2B illustrates the workflow for the two loss functions, in which the target prediction loss is shown as multiple heads. Fig. 2C shows more details on the multi-headed binary classification model.

However, to combine $J_x$ and $J_y$, it is not acceptable to simply add them up because they have different scales. [17] proposed a method to minimize multi-task loss functions in the same time. The authors showed that maximizing a multi-task likelihood is equivalent to minimizing a combination of individual task losses with variance dependent weights, with an assumption of homoscedastic uncertainty for each task. They achieved better performance with the multi-task weightings than individual tasks trained separately. Applying their method to our model, we have the loss function as Equation 9 below, with $\sigma_1$ and $\sigma_2$ standing for trainable variance parameters for $J_x$ from Equation 1 and $J_y$ from Equation 7:

$$J = \frac{1}{2\sigma_1^2} J_x + \frac{1}{\sigma_2^2} J_y + log\sigma_1 + log\sigma_2 \quad (9)$$

### 2.3 The Stacked Model

We used the manufacturing data from Seagate, specifically sensors captured during the wafer processing steps within one family of deposition tools. This data has hundreds of dimensions. To make the model powerful enough to deal with the complicated data sets, a stacked version of the variance weighted multi-headed autoencoder based classifier is developed, shown as Fig. 3.

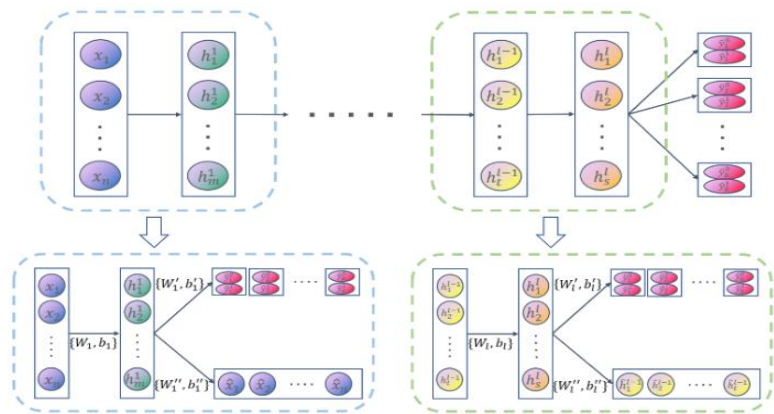

Fig. 3. Greedy layer-wise training with a multi-headed classifier

As discussed above, the auto-encoder part of this model is encoding the input data as a lower-dimension representation and reconstructing both the input layer and the target variables. Here the target variables include multiple heads, each head is a binary classification. The loss function for this model contains two parts: one is the auto-encoder loss functions, and another is the multi-binary-classification loss function, as in Equation 9.

This model adds a multi-headed classifier to auto-encoder, making it able to learn a representation that captures the features that are most relevant to characterizing and predicting a series of key wafer quality-control measurements. Given that the dimension of the heads is much smaller than the features, this model only adds negligible complexity to auto-encoder thus trains as fast as traditional autoencoders. On the other hand, the vectorization of the multiple heads also accelerates the calculation.

## 2.4 Model Implementation

The training process is illustrated in Table. 1, given a high dimensional input, a stacked encoder-decoder structure is used to reduce the dimension gradually. While the input data dimension is 632, and output dimension is 16, we chose four layers of the stacked encoder-decoder structure, with the layers as [400, 200, 100, 50] so that in each iteration the dimension is effectively reduced and not too much information is lost.

Table. 1 VWMHQAE pseudo code

| **Training process** |
| --- |
| > Decide the structure of the model with number of layers n and each layer $[l_1\text{-}l_n]$, and input data $X_0$, output data $y$ |
| > Loop: <br> for k in 1 to n: <br> - Train a layer of the encoder-decoder with input $X_{k-1}$ and output $(X_{k-1}, y)$, and get the hidden layer $X_k$ with dimension $l_k$ |
| > Based on the final layer of encoder-decoder, extract the hidden layer $X_n$, and train a logistic regression classifier with output $y$, get prediction $\hat{y}$ |
| > Evaluate the results by comparing $y$ and the predicted $\hat{y}$: <br> For each output $y_i$ in $y$: <br> - Count the correctly predicted positive/negative cases and calculate the recall and accuracy |

To further illustrate the effectiveness of the multi-heading mechanism, Fig. 4 shows the comparison of the multi-headed model with the single-head model which taking the heads as a categorical variable instead. In the original data, the feature indicating which measurement is done on that wafer is a column that has no difference with any other categorical variables. To what we know, researchers have been taking this kind of variables as categorical variables in the input. However, we found that the measurement feature can not only be a categorical variable in input data, but also be an output head given that the distinct values are not too many (8 in this data set). As shown in Fig. 4, it turns out that taking the measurement variable as output heads not only improves the interpretability, but also helps the model converges much faster and with less fluctuations in the loss function as in Equation. 9. Noted that the loss values went to negative because of the variance weighting in the formula, and the absolute values in Fig. 4 may not linearly reflect the accuracy of the model. However, it's shown that the multi-headed mechanism accelerates the convergence and leads to a lower value of the loss function.

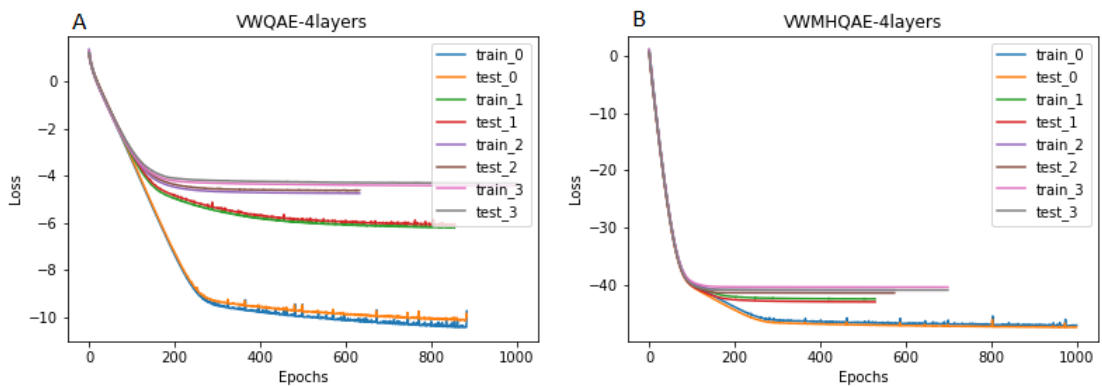

Fig. 4 Comparison for the multi-headed model with a model taking the heads as categorical feature. A) Variance-weighted quality-driven autoencoder model, treating measurement feature as input categorical variable, B) Variance-weighted multi-headed quality-driven autoencoder model, treating measurement feature as output heads.

In the following case study, we used a four-layer stacked model to learn on a data set with 632 dimensions which has ~0.2 million data points. We used an Adam optimizer with learning rate as 1e-3, batch size as 512, and an early-stopping with minimum delta as 1e-5. There are 8 different measurement steps, corresponding to 16 heads in the model. With the dimensions of the four hidden layers as (400, 200, 100, 50), the total number of parameters in the model is shown in Table. 2:

Table. 2. Parameters for the 4-layer VWMHQAE model. (a) encoder parameters, (b) reconstruction (decoder x) parameters, (c) prediction (decoder y) and the final-layer classifier parameters.

| Layers | Number of parameters |
| --- | --- |
| Autoencoder_1 | $633 * 400^a + 401 * 632^b + 401 * 16^c$ |
| Autoencoder_2 | $401 * 200^a + 201 * 400^b + 201 * 16^c$ |
| Autoencoder_3 | $201 * 100^a + 101 * 200^b + 101 * 16^c$ |
| Autoencoder_4 | $101 * 50^a + 51 * 100^b + 51 * 16^c$ |
| Classifier | $51 * 16^c$ |
| Total | 730562 |

Comparing with a regular autoencoder model with the same layers, our VWMHQAE model only has 1.7% more parameters. The complexity of our model is almost the same as a stacked autoencoder. With a training data of ~0.16 million and dimension 632, the training for a four-layer VWMHQAE model takes about 30 minutes on a NVIDIA Tesla V100 SXM2 GPU.

## 3    Case Study

The main purpose of the work is to find a generalized and computationally feasible soft sensor technique that can be applied to a diverse family of equipment. The model also needs to perform well in class imbalance and be able to simultaneously learn to predict output of all multiple measurement steps that are critical to ensure the quality of a particular point.

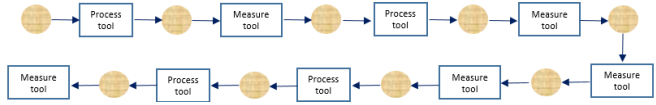

Fig. 5. Simplified wafer manufacturing process showing a wafer moving through the line getting processed and measured

In a complex manufacturing setting as in Fig. 5, effective process requires that we have measurements in place for our processes that can be monitored and controlled automatically. While these measurements are critical to assess the quality of the processes, they also add non-trivial cycle time into the manufacturing process. An optimal trade-off needs to be maintained to ensure the right balance of quality and time to market. A soft sensing methodology that maps the abundant process variables to quality metrics with high degree of recall could potentially be exploited to reduce the time to ship by enabling such processes to be skipped based on the process variable readings. In the semiconductor industry, this method is widely referred to as virtual metrology.

In complicated fabricating plant, measurement steps can comprise 20-40% of the content. The more complicated a process gets, the more frequently and exhaustively that process is measured. Fig. 5 shows how simplified version of how a wafer can go from a process tool to a measurement tool and then to another process tool or measurement tool and so on. The main process steps involve deposition of metal layers and dielectric/insulating materials. The deposition process consists of depositing a material on the wafer through several technologies: PVD (Physical Vapor Deposition), PECVD (Plasma Enhanced Chemical Vapor Deposition), SACVD (Sub-Atmospheric Chemical Vapor Deposition), LPCVD (Low Pressure Chemical Vapor Deposition) or more recently ALD (Atomic Layer Deposition).

### 3.1 Data Description

For the purpose of this work, one year's worth of process variables (sensor) data was used, with the corresponding measurement (quality) variables. The sampling rate of the sensors is mostly per second. The measurement data is collected as a wafer is physically measured which could range from minutes to hours. The data is published at https://github.com/Seagate/softsensing_data.

The process tool used in this work is a vacuum tool that deposits thin film metal as in Fig. 6. Processes on this tool include both etching and deposition steps. Materials are sputtered from the target deposits on the substrate to create a film. In combination with the substrate fixture that rotates and changes the angle, this type of tooling system also delivers a broad range of control over sidewall coatings, trench filling and liftoff profiles.

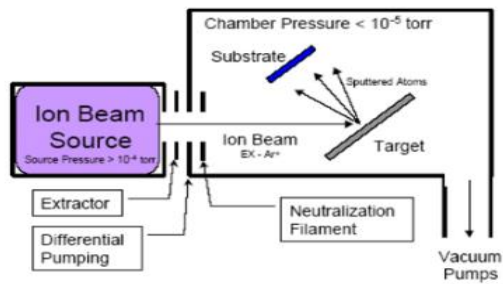

Fig. 6 $P_1$ tool process flow

The tool family considered in this study comprises 14 physical deposition equipment. Each equipment has 130 sensors installed which are captured at a frequency of about every second. The proposed model uses the aggregated statistic from the raw observations as in Table. 3.

Table. 3 Statistics on the process variables (sensors)

| Tool family | Tools | Sensors | Observations |
|---|---|---|---|
| $P_1$ | 14 | 130 | 329,506,677 |

The critical parameters measured for this family of tools are magnetics, thickness, composition, resistivity, density, and roughness. These parameters are critical to the quality of the wafers and are measured by a family of high precision measurement equipment that come high capital costs. As shown in Fig. 5, a wafer can go through one or more measurement tools, dependent on the criticality and complexity of the process that needs to be measured. The sample size of measurement variables is much lower than process variables. A measurement taken at any physical measurement tool comprises a step. A wafer coming out of a tool for example, from a family of $P_1$, could possibly have 8 measurement steps. Each measurement step is regarded as an individual task in this paper, referred as $Y_1 - Y_8$. The measurements examine different quality parameters of the wafer including thickness, magnetics, resistivity, composition, etc., and the number of data samples vary from ~2,000 to ~25,000, and the imbalance ratios range from 2 to 225.

### 3.2 Data Imbalance

As is expected from most real-world data, the sensor data used here are highly imbalanced. Data imbalance in this study is considered from two perspectives. First, there is class imbalance, which is the ratio of positive versus negative classes. We refer to the difference in scale or unit between the mean squared and the crossentropy losses used in the combined loss function as imbalance in the task. The class imbalance biases the learning towards the majority class due to its increased prior probability. The imbalance, or the difference in units of the combined loss function used in the two tasks of the proposed model can result in imbalanced weights for the different losses, biasing the learning towards the higher weighted loss and also possibly causing the lower weighted loss to suffer from vanishing gradients.

Despite recent advances and successes of deep learning, there has not been enough empirical work in problems where high class imbalance exists [20].

Methods for handling class imbalance in machine learning fall into 3 categories: data-level techniques, algorithm-level methods, and hybrid approaches. [18]

Data level techniques address the issue of class imbalance through data sampling methods. Algorithm-level techniques are usually incorporated in the cost function, so the learner attempts to mitigate the bias towards the majority class during learning. Hybrid approaches combine the two.

Methods for handling the imbalance in the units or the weights of different losses in a combined loss function involve using a naïve weighted sum of losses [19]. Kendal et al use a Bayesian approach to learn multiple objectives in a loss function [17].

In order to determine the most reasonable fit for the proposed soft sensing architecture, we experimented with various techniques, the results are discussed in the section below.

## 4 Results and Discussion

Both the data level and algorithm level techniques were tested to address the class imbalance in this work. First, synthetic minority over-sampling technique (SMOTE) was utilized since it is a widely accepted benchmark for learning from imbalanced data [21]. This method utilizes the oversampling approach to rebalance the original training set. Instead of applying a simple replication of the minority class instances, it introduces synthetic examples which are created by interpolation between several minority class instances within a defined neighborhood.

For soft sensor modeling, however, algorithm-level approaches were shown to be more effective. For example, the loss function was adjusted to take weights into consideration in to reduce bias towards the negative class. [Eq. 7]. The introduction of the weights in the loss function allows the minority samples to contribute more to the lost. The successful use of class weight was also demonstrated by Wang et al. [22] and Lin et al. [23].

For evaluation and comparison purposes, we consider true positive rate(recall) to be a more useful metric because the cost of false negative is much higher than the cost of false positive in this case. An additional $F_\beta$ metric is also used to evaluate and compare the models. $F_\beta$ is a generalized version of F score to show the emphasis on recall, proposed by [24] and further clarified by [25].

The idea is to give a weight to recall and precision in the F-score formula:

$$F_\beta = \frac{(1+\beta^2)*P*R}{\beta^2*P+R} \tag{10}$$

We follow the practice from [20], and use the imbalance ratio for $\beta$ to compare experimental results across all models. The imbalance ratio is:

$$\beta = \frac{max_i\{|C_i|\}}{min_i\{|C_i|\}} \tag{11}$$

where $C_i$ is the number of observations in class $i$, $max_i\{|C_i|\}$ and $min_i\{|C_i|\}$ return the maximum and minimum class size over all i classes, respectively.

We tested the proposed architecture using several data imbalance techniques to determine what best suits the data.

Both SMOTE and Weighted Class techniques were tested on models which included logistic regression (LR), neural network (NN) with three fully connected layers, QAE+LR/NN (LR/NN using quality-based latent variables) and SQAE (LR/NN using quality-based stacked latent variables). Table. 4 shows that algorithm-based approach, in this case, weighted class outperforms SMOTE in 6 of the 8 outputs. Therefore, the subsequent experiments use class weights to handle class imbalance.

Additionally, two different approaches were used in order to handle the different scales in the combined loss of the models. In the first approach, we use a naïve weighted sum of losses to combine the reconstruction loss and the binary cross entropy loss. The second approach sees the two loss functions as a multi-task learning problem and uses a variance based weighting mechanism as proposed by [15].

Table. 4 Imbalance performance compare across outputs

| Output | IMB Method | Recall | $F_{\beta\_imb}$ |
|---|---|---|---|
| $Y_1$ | SMOTE | 0.5277 | 0.5275 |
| | WEIGHTED CLASS | **0.7447** | **0.7441** |
| $Y_2$ | SMOTE | 0.4627 | 0.4615 |
| | WEIGHTED CLASS | **0.5608** | **0.5583** |
| $Y_3$ | SMOTE | 0.8940 | 0.8917 |
| | WEIGHTED CLASS | **0.9333** | **0.9292** |
| $Y_4$ | SMOTE | **0.7801** | **0.7776** |
| | WEIGHTED CLASS | 0.7673 | 0.7642 |
| $Y_5$ | SMOTE | 0.7876 | 0.7836 |
| | WEIGHTED CLASS | **0.8371** | **0.8305** |
| $Y_6$ | SMOTE | **0.7500** | **0.6826** |
| | WEIGHTED CLASS | 0.6494 | 0.6189 |
| $Y_7$ | SMOTE | 0.5208 | 0.5198 |
| | WEIGHTED CLASS | **0.6015** | **0.6000** |
| $Y_8$ | SMOTE | 0.5330 | 0.5317 |
| | WEIGHTED CLASS | **0.6659** | **0.6637** |

As shown in Table. 5, results of quality driven autoencoder versus the stacked quality driven autoencoder by each output. On average, we found no significant performance gain with stacked autoencoders. However, as Fig. 7 shows, autoencoder based models consistently perform better than models without pre-trained autoencoder. The highest performing model across all outputs is a weighted class and variance weighted stacked autoencoder with a logistic regression classifier (recall=0.75/$F_\beta$=0.74), followed very closely by naively weighted quality autoencoder model (recall=0.74/$F_\beta$=0.73).

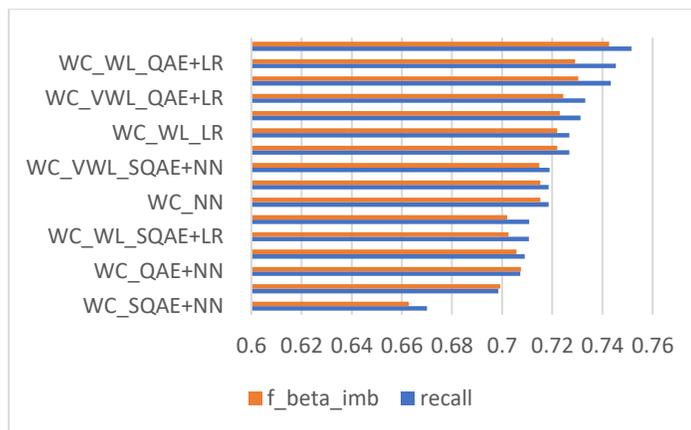

Fig. 7 Average Output Performance for variations of the VWMHQAE model. The abbreviations are: AE: autoencoder, Q: quality-driven, S: stacked (with hidden layers [400,200,100,50]), WC: weighted class, WL: weighted loss, VWL: variance weighted loss, LR: linear regression classifier, NN: fully connected neural net classifier (with hidden layers [100,50])

Table. 5 Autoencoder vs. stacked autoencoder compare across outputs. QAE stands for a single layer model, and SQAE for a 4-layer model

| Output | AE | Recall | $F_{\beta\_imb}$ |
|---|---|---|---|
| $Y_1$ | QAE | **0.8333** | 0.8323 |
|  | SQAE | **0.8333** | **0.8324** |
| $Y_2$ | QAE | **0.5263** | **0.5242** |
|  | SQAE | 0.50263 | 0.5008 |
| $Y_3$ | QAE | 0.9106 | 0.9074 |
|  | SQAE | **0.9349** | **0.9313** |
| $Y_4$ | QAE | **0.7881** | **0.7853** |
|  | SQAE | 0.7711 | 0.76802 |
| $Y_5$ | QAE | 0.8226 | 0.8170 |
|  | SQAE | **0.8231** | **0.8171** |
| $Y_6$ | QAE | **0.7089** | **0.6573** |
|  | SQAE | 0.7059 | 0.6481 |
| $Y_7$ | QAE | 0.5656 | 0.5643 |
|  | SQAE | **0.5906** | **0.5891** |
| $Y_8$ | QAE | **0.6206** | **0.6189** |
|  | SQAE | 0.5531 | 0.5516 |

## 5    Conclusion and Future Work

In this work, we present an efficient multi-headed autoencoder based architecture that learns to simultaneously classify multiple outcomes by exploiting the feature representation guided by quality relevant features. We consider this to be an important contribution to the operationalization of soft sensor modeling in a manufacturing environment with complex non-linear interactions among thousands of complex operations. The capability of using sensors to characterize impact on the outcome of multiple steps in a feed forward manner allows us to be more proactive about the potential build issues that may arise, and at the same time enables skipping of certain measurement steps, thereby saving tool capacity and improving time to ship window.

We used several techniques to address the challenge of class imbalance and the imbalance in the scale of the different loss function and showed that in this dataset, algorithm-based techniques work better.

This architecture can be trivially generalized to regression models or multi-class classification models.

Future work can extend the architecture to include the output driven loss functions at each layer of the stacked autoencoder instead of introducing it at the top layer only. Convolutional network can also be applied on the raw time-series sensor data to extract exploit the locality and stationarity features in the data. We also believe that the generality of this architecture can further be improved by exercising these soft sensing models on a diverse set of equipment with their own function specific sensors. In the meantime, we have the world's largest manufacturing and sensing data sets from Seagate Technology, and we hope to open-source these data sets for the entire community to develop more scalable and applicable AI techniques.